\documentclass[preprintnumbers,twocolumn,floats,aps,nofootinbib]{revtex4-1}
\pdfoutput=1
\usepackage{graphicx}
\usepackage{color, graphics}
\usepackage{amsmath,amssymb,amsfonts}
\usepackage{verbatim}   
\usepackage{subfigure}  
\usepackage{acronym}
\usepackage{hyperref}
\usepackage{mathrsfs}
\usepackage{multirow}
 \usepackage{cancel}
 \usepackage{url}
  \usepackage{slashed}
 \usepackage{hyperref}

\def\beq{\begin{equation}\begin{aligned}}
\def\eeq{\end{aligned}\end{equation}}
\def\OO{\mathcal{O}}

\begin{document}

\title{On Baryogenesis from a Complex Inflaton}

\author{James Unwin}
\email{junwin@nd.edu}
\affiliation{Department of Physics, University of Notre Dame,  Notre Dame, IN, 46556, USA}

\begin{abstract} 

We derive the particle asymmetry due to inflationary baryogenesis involving a complex inflaton,  obtaining a different result to that in the literature. While asymmetries are found to be significantly smaller than previously calculated, in certain parameter regions baryogenesis can still be achieved.

\end{abstract}

\maketitle

\section{Introduction}

Whilst the general principles behind the generation of particle-antiparticle asymmetries are well understood \cite{Sakharov:1967dj}, the specific mechanism through which baryogenesis occurred remains a mystery. What is apparent is that the Standard Model alone does not give rise to the appropriate conditions to realise baryogenesis via the electroweak phase transition, see e.g.~\cite{Morrissey:2012db}. A particularly interesting prospect is that the baryon asymmetry might be generated due to inflationary physics \cite{Delepine:2006rn,Hook:2014mla,Alexander:2004us,Affleck:1984fy,Hertzberg:2013mba,Hertzberg:2013jba,BasteroGil:2011cx}. 

In this letter we re-examine the inflationary baryogenesis scenario proposed recently in \cite{Hertzberg:2013mba,Hertzberg:2013jba}. We rederive the parametric form of the asymmetry, finding a significantly different result. Subsequently, we identify the parameter regions which are simultaneously consistent with the cosmological evidence for inflation \cite{Hinshaw:2012aka,Ade:2015lrj} and allow for successful baryogenesis. Moreover, we discuss issues related to effective field theory intuition and procedures for changing between dimensionful and dimensionless sets of variables, which are more generally applicable.

Hertzberg and Karouby \cite{Hertzberg:2013mba,Hertzberg:2013jba} considered the possibility that the inflaton was a complex field
\beq
\phi(t)=\frac{1}{\sqrt{2}}\rho(t)e^{i\theta(t)}~,
\label{phi}
\eeq
which carries a conserved\footnote{Up to Planck scale, $M_{\rm Pl}$, effects which are expected to violate all continuous global symmetries.} global quantum number. The requirements for generating a particle asymmetry in a given global charge is a period of out-of-equilibrium dynamics, together with violation of $C$, $CP$, and the associated global symmetry  \cite{Sakharov:1967dj}. In the scenario at hand, the out-of-equilibrium dynamics is driven by inflation. Further, $C$ and $CP$ can be broken spontaneously due to the  initial phase of the inflaton field $\theta_i$ (reminiscent of the Affleck-Dine mechanism \cite{Affleck:1984fy}). Finally, the violation of the inflaton global symmetry is sourced from small breaking terms in the potential. We shall take a simple quadratic potential (as used commonly  in chaotic inflation  \cite{Linde:1983gd}) supplemented by a single dimension-$n$ operator (for $n\geq3$) which breaks the U(1) global symmetry and, following  \cite{Hertzberg:2013mba}, we assume a potential of the form
\beq
V(\phi,\phi^*)=\frac{1}{2}m^2|\phi|^2+\lambda\left(\frac{1}{\Lambda}\right)^{n-4}(\phi^n+\phi^*{}^n)~,
\label{V}
\eeq
where $\lambda$ is a dimensionless coupling, $\Lambda$ has mass dimension one (deviating from the notation of \cite{Hertzberg:2013mba}). Note that as the latter term causes a perturbation from the  quadratic potential, its magnitude can be constrained by cosmological observations, as we shall discuss.

A natural measure of particle asymmetries is
\beq
A_\infty\equiv\frac{n-\bar n}{n+\bar n}~,
\label{A}
\eeq 
in terms of the number densities of particles $n$ and antiparticles $\bar n$. This quantity is bounded by
\beq 
0\leq |A_\infty|\leq1~.
\label{bound}
\eeq
Extremal values correspond to equal populations $n=\bar n$ for $A_\infty=0$ or a completely asymmetric population with vanishing $n$ or $\bar n$ for $|A_\infty|=1$. Once an asymmetry is established in the inflaton charge, the inflaton can decay in a manner which transfers the asymmetry to baryons. It has been suggested that an appropriate measure of the inflaton asymmetry at early time is
\beq
A_0= \frac{m(n-\bar n)}{\epsilon}~,
\label{A01}
\eeq
where $\epsilon$ is the energy density and the subscripts $0$ and $\infty$ distinguish the  asymmetry at early and late time.  At late times the energy density is determined by the non-relativistic gas of $\phi$ and $\phi^*$, thus $\epsilon=m(n+\bar n)$ and the asymmetry reduces to the familiar form of eq.~(\ref{A}). 

It was argued in \cite{Hertzberg:2013mba} that, by evaluating eq.~(\ref{A01}), the late time asymmetry can be expressed in terms of fundamental quantities as follows
\beq
A_\infty^{\rm (HK)}\sim
-c_n\lambda
\left( \frac{M_{\rm Pl}^{n-2}}{m^2\Lambda^{n-4}}\right)\sin(\theta_i n)~,
 \label{HK}
\eeq 
where $c_n$ is a constant. The value of $c_n$ for $3\leq n \leq10$ is calculated in \cite{Hertzberg:2013mba}, and the first few are quoted below
\beq
c_3\approx7, \quad c_4\approx 11.5, \quad c_5\approx 14.4, \quad c_6\approx 21.8~.
\label{cn}
\eeq
Whilst the prospect of generating baryogenesis through the dynamics of a complex inflaton is rather elegant, the form of $A^{\rm (HK)}_\infty$ raises some questions, in particular eq.~(\ref{HK}) seemingly permits values for the asymmetry greater than unity. Consider, for instance,  $n=5$ with $\lambda=1$ and $\sin(5\theta_i)=-1$, taking motivated parameter values gives
\beq
A^{\rm (HK)}_\infty\sim
10^{13}~\left(\frac{10^{16}~{\rm GeV}}{\Lambda}\right)\left(\frac{10^{13}~{\rm GeV}}{m}\right)^2~.
\eeq
This is seemingly in contradiction with eq.~(\ref{bound}) and the definition of the asymmetry.
Additionally, the scaling behaviour of $A^{\rm (HK)}_\infty$ is counterintuitive, as it does not follow our expectations from effective field theory. The asymmetry receives, unbounded, sequentially larger contributions from operators with increasing mass dimension. 

The purpose of this paper is to rederive the form of the asymmetry due to a complex inflaton $A_\infty$. We obtain a result which satisfies $|A_\infty|\leq1$ and effective field theory considerations, but differs significantly in form from eq.~(\ref{HK}). In the Appendix we give a careful account of the differences between the derivation here and that of \cite{Hertzberg:2013mba}, along with arguments in favour of our approach.

\section{The inflaton asymmetry}

The asymmetry $A_0$ can be expressed in terms of the overall charge difference $\Delta N_\phi$, which is related to the particle asymmetry per comoving volume $n-\bar n$ as follows
\beq
A_0= \frac{m(n-\bar n)}{\epsilon} = \frac{m}{\epsilon} \left(\frac{\Delta N_\phi}{V_{\rm co}a^3}\right)~,
\label{A0}
\eeq
where $V_{\rm co}$ is the comoving volume, and  $a(t)$ is the scale factor.
By examining the evolution of the number of inflatons $N_\phi$ relative to the number of anti-inflatons $N_{\bar\phi}$ one obtains
\beq
\Delta N_\phi\equiv N_\phi-N_{\bar\phi}=iV_{\rm co} a^3(\phi^*\dot\phi-\dot\phi^*\phi)~,
\label{eq1}
\eeq
where we have assumed an FRW metric. The equation of motion (EoM) for $\phi$ is given by
\beq
\ddot{\phi}+3H \dot{\phi}+m^2\phi+\lambda n\left(\frac{1}{\Lambda}\right)^{n-4} \phi^*{}^{n-1}=0~,
\eeq
where $H=\dot a/a$ is the Hubble parameter. The final charge difference $\Delta N_\phi(t_f)$ can be found by taking the time derivative of eq.~(\ref{eq1}) and using the EoM, to obtain
\begin{align*}
\Delta N_\phi(t_f)
\simeq&~\Delta N_\phi(t_i)\\&+i\lambda \left(\frac{1}{\Lambda}\right)^{n-4}n V_{\rm co} \int_{t_i}^{t_f}a(t)^3\left[\phi(t)^n-\phi^*(t)^n\right].
\end{align*}
Moreover, as any initial asymmetry is likely erased via inflation: $\Delta N_\phi(t_i)\simeq0$. Thus, to $\OO(\lambda)$ this gives
\beq
\Delta N_\phi(t_f)
\simeq -\left(\frac{1}{\Lambda}\right)^{n-4}\frac{\lambda nV_{\rm co} }{2^{\frac{n}{2}-1}} \int_{t_i}^{t_f}a(t)^3\rho(t)^n\sin(n\theta(t)).
\eeq
Since at zeroth order in $\lambda$ the argument does not evolve, we can take $\theta(t)=\theta_i$. 
Moreover, at zeroth order in $\lambda$, the radial component $\rho$ is a real valued function satisfying
\beq
\ddot \rho+3H\dot\rho+m^2\rho=0~,
\label{er}
\eeq
with the associated Friedmann equation
\beq
H^2=\frac{1}{6M_{\rm Pl}^2}\left(\dot\rho^2+m^2\rho^2\right)
\equiv\frac{\epsilon}{3M_{\rm Pl}^2}
~.
\label{eps}
\eeq
Then, working at lowest order, one can express $\Delta N_\phi(t_f)$ in terms of the radial component to obtain \cite{Hertzberg:2013mba} 
\beq
\Delta N_\phi(t_f)\simeq-\lambda\left(\frac{1}{\Lambda}\right)^{n-4} \frac{V_{\rm co} n}{2^{\frac{n}{2}-1}}~\sin(\theta_i n)~I(t_i,t_f)~,
\label{DN}
\eeq 
with 
\beq
I(t_i,t_f)=\int_{t_i}^{t_f} {\rm d}t~ a(t)^3\rho(t)^n~.
\label{ei}
\eeq 
It follows that the asymmetry can be expressed as follows
\beq
A_0=
-\frac{m}{a^3\epsilon} 
\left(\frac{1}{\Lambda}\right)^{n-4} \lambda\frac{n}{2^{\frac{n}{2}-1}}\sin(\theta_i n)~I(t_i,t_f)~.
\label{start}
\eeq

Here is where our derivation differs crucially from \cite{Hertzberg:2013mba} (see Appendix). We make a change of variables such that everything is measured in units of inflaton mass $m$ 
\beq
\tau\equiv mt~,
\qquad
\hat\rho\equiv\frac{\rho}{m}~,
\qquad
\hat H\equiv \frac{H}{m}~.
\label{scale}
\eeq
Thus each of the rescaled quantities is dimensionless (reminiscent of the $M_{\rm Pl}$-units sometimes employed).   The scaling leads to a dimensionless version of  eq.~(\ref{er})
\beq
\frac{\rm d^2}{{\rm d}\tau^2}{\hat\rho}+3\hat H\frac{\rm d}{{\rm d}\tau}{\hat\rho}+\hat\rho=0~.
\label{diml}
\eeq
The corresponding dimensionless Friedmann equation can be written in terms of $\hat \epsilon\equiv\epsilon/(m M_{\rm Pl})^2$
\beq
\hat H^2
=\frac{1}{6m^2M_{\rm Pl}^2}\left(\left[\frac{\rm d \hat\rho}{{\rm d}\tau}\right]^2+\hat\rho^2\right)
=\frac{\hat\epsilon}{3}~.
\label{bare}
\eeq
Following the notation of \cite{Hertzberg:2013mba}, we introduce 
\beq
f_n&= \frac{n}{2^{\frac{n}{2}-1}} \frac{1}{a^3\hat\epsilon}~ \bar I(\tau_i,\tau_f)~,
\label{fn}
\eeq
in terms of the scaled quantity
\beq
\bar I(\tau_i,\tau_f)=\int_{\tau_i}^{\tau_f} {\rm d}\tau~ a(\tau)^3\rho(\tau)^n~.
\label{eib}
\eeq
It follows that eq.~(\ref{start})  can be rewritten as
\beq
A_0=-\lambda f_n \left(\frac{m^{n-2}}{M_{\rm Pl}^2\Lambda^{n-4}}\right)\sin(\theta_i n)~.
\label{AA}
\eeq
To obtain the late time asymmetry we should evaluate $f_n$ in the limit $\tau_i,\tau_f\rightarrow\pm\infty$. Then making the replacement $f_n\rightarrow c_n=f_n(\tau_i\rightarrow-\infty,\tau_f\rightarrow\infty)$ gives our main result
\beq
A_\infty=-c_n\lambda  
\left(\frac{m^{n-2}}{M_{\rm Pl}^2\Lambda^{n-4}}\right)\sin(\theta_i n)~,
\label{asym}
\eeq
where the values of $c_n$ are as in eq.~(\ref{cn}). Observe that $A_\infty$ above is distinct from $A_\infty^{\rm (HK)}$ of \cite{Hertzberg:2013mba} (quoted in eq.~(\ref{HK})). In particular, note that for all $n$ the asymmetry is bounded $|A_\infty|\leq1$, and contributions are suppressed for increasing $n$, as expected from effective field theory considerations.

\section{The Baryon asymmetry}

The asymmetry in baryons is often defined as follows
\beq
\eta_b=\frac{n_b-\bar n_{b}}{n_\gamma}\simeq6\times10^{-10}~.
\eeq
This is similar to eq.~(\ref{A}), but here the difference between baryons $n_b$ and anti-baryons $\bar n_{ b}$ is normalised relative to the photon number density $n_\gamma$. In the previous section we  derived the parametric form of the asymmetry in the inflaton global charge $A_\infty$, which is related to  $\eta_b$ via \cite{Hertzberg:2013mba}
\beq 
\eta_b&
\sim
g_*^{\frac{3}{4}} A_\infty\left(\frac{\sqrt{\Gamma M_{\rm Pl}}}{m}\right)~.
\eeq
We calculate here the magnitude of $\eta_b$ which arises due to the inflaton asymmetry and identify parameter regions in which the observed baryon asymmetry can be realised.

As only quarks carry baryon number in the Standard Model, the first gauge and Lorentz invariant baryon number violating operator is $\phi QQQL$, however mild extensions of the Standard Model can alter this. Incidentally, leptogenesis might be accomplished with lower dimension operators. We will not pursuit these possibilities further here, but shall return to them in \cite{companion}.
 For our purposes we shall simply suppose that the inflaton decays dominantly via a dimension-$p$ operator, suppressed by a scale $M$. Thus the decay rate is parametrically
\beq
\Gamma_\phi\sim m\left(\frac{m}{M}\right)^{2(p-4)}~.
\eeq 
It would be quite natural to identify $M$ with $\Lambda$, but for the moment we shall maintain the more general possibility that these scales are distinguished. Substituting the forms of $\Gamma_\phi$ and $A_\infty$, and assuming that each inflaton decay violates baryon number by one unit (as with $\phi QQQL$), the resultant baryon asymmetry is given by
\beq
\eta_b&
\sim
- c_n \lambda
g_*^{\frac{3}{4}} 
\left(\frac{m}{M}\right)^{p-4}
\left(\frac{m}{\Lambda}\right)^{n-4}
\left(\frac{m}{M_{\rm Pl}}\right)^{\frac{3}{2}}
\sin(\theta_i n).
\eeq

Further, up to a dependance on the number of e-folds of inflation $N$, the observed value \cite{Hinshaw:2012aka} of the squared amplitude of density fluctuations $\Delta_R^2\approx2.45\times10^{-9}$ fixes  the inflaton mass in models of single field slow roll inflation
\beq
m\simeq 
\frac{\sqrt{6}\pi\Delta_R M_{\rm Pl}}{N}
\simeq 1.5\times10^{13}~{\rm GeV}\left(\frac{60}{N}\right)~.
\eeq
Notably, the observed baryon asymmetry can be readily reproduced with this value of inflaton mass. For a dimension five breaking operator ($n=5$) and a dimension seven transfer operator ($p=7$) generated at the scale $\Lambda=M$, the observed $\eta_b$ can be achieved as below
\beq
\eta_b\sim 
10^{-9}&
\left(\frac{\lambda}{1}\right)
 \left(\frac{m}{10^{13}{\rm GeV}}\right)^{\frac{11}{2}}
 \left(\frac{10^{14}{\rm GeV}}{\Lambda}\right)^{4}
\left(\frac{\sin5\theta_i}{1}\right),
\label{no}
\eeq
where we have taken $g_*\simeq100$. 

\pagebreak

In \cite{Hertzberg:2013mba} it is argued, using $A^{\rm (HK)}_\infty$, that $\eta_b\sim10^{-9}$ can be obtained for $M=\Lambda\sim10^{16}{\rm GeV}$. As a result decays via $\phi QQQL$ are subdominant to dimension five U(1)-violating $M_{\rm Pl}$-suppressed decays and the asymmetry is erased unless some symmetry forbids these operators. However, using instead the form of $A_\infty$ from eq.(\ref{asym}), such {\em ad hoc} symmetries are no longer necessary.\footnote{Further details and discussion will be presented in \cite{companion}.}

We conclude that realistic values of the baryon asymmetry can in principle be generated. However, thus far we have paid no heed to restrictions from inflation, which we examine next. To reproduce the predictions of chaotic inflation it is required that the U(1) breaking term, which perturbs the quadratic potential, is small \cite{Hertzberg:2013mba}
\beq 
\left(\frac{\lambda}{2^{\frac{n}{2}-1}\Lambda^{n-4}}\right)\rho_i^n\cos(n\theta_i)\ll\frac{1}{2}m^2\rho_i^2~.
\label{31}
\eeq
Since one typically expects $\rho_i\sim M_{\rm Pl}$, it follows that
\beq
\left(\frac{\lambda}{2^{\frac{n}{2}-2}}\right)\cos(n\theta_i)\ll \frac{m^2\Lambda^{n-4}}{M_{\rm Pl}^{n-2}}\ll1~.
\label{cons}
\eeq 
For general $\theta_i$ this constraint is highly problematic, as it implies $\lambda\ll1$; 
e.g.~with $\sin(n\theta_i)\sim\cos(n\theta_i)\sim\frac{1}{\sqrt{2}}$ and $n=5$
 to avoid disturbing the quadratic potential requires 
 \beq
\lambda\ll\frac{\Lambda m^2}{M_{\rm Pl}^{3}}
\simeq 10^{-14} 
\left(\frac{m}{10^{13}~{\rm Gev}}\right)^{2}
 \left(\frac{\Lambda}{10^{14}~{\rm Gev}}\right)~.
\eeq
Such values of $\lambda$ are typically too small to realise the observed baryon asymmetry. In the example studied in eq.~(\ref{no}), this leads to baryon asymmetries $\eta_b\ll10^{-23}$.

However, observe that eq.~(\ref{cons}) is trivially satisfied for $\cos(n\theta_i)\approx0$ (also note that in this case $|A_\infty|$ is maximal, as $\sin(n\theta_i)\approx\pm1$). Thus, for special values\footnote{The forms of eq.~(\ref{asym}) \& eq.~(\ref{31}) can vary if the symmetry violating operator is changed (e.g.~$\Lambda^{4-n}\phi^{n-1}\phi^*+{\rm c.c.}$). Thus so can the values of $\theta_i$ for which eq.~(\ref{31}) is automatically satisfied.${}^2$}
of $\theta_i$ inflationary cosmology is unperturbed. It would be interesting to investigate whether there are mechanisms which can fix $\theta_i$ at these distinguished values. From an alternative perspective, given that prior to inflation the field $\phi$ takes different values of $\theta_i$ in different local patches, one of which subsequently inflates to form the visible universe, this might allow for an anthropic explanation.

\section{Conclusions}

Our main result is the expression of $A_\infty$, the magnitude of the particle asymmetry expected due to a complex inflaton, given in eq.~(\ref{asym}). The form of the asymmetry is characteristically different from $A_\infty^{\rm (HK)}$ derived in \cite{Hertzberg:2013mba,Hertzberg:2013jba}, which we quote in eq.~(\ref{HK}).
Arguments were presented for why we believe the asymmetry derived here to be correct (and the Appendix explains the source of this deviation). 

Using the form of $A_\infty$ derived in eq.~(\ref{asym}), we identified  parameter regions in which an appropriate baryon asymmetry can be generated without perturbing the quadratic potential which drives inflation. In particular, we argued that for the simple models studied, inflationary cosmology and the observed baryon asymmetry can only be simultaneously reproduced for special values of $\theta_i$.

The calculations presented here have involved purely perturbative inflationary processes, however there have been some efforts \cite{Lozanov:2014zfa,Hertzberg:2014jza} to examine analogous baryogenesis scenarios involving preheating \cite{Kofman:1994rk,Traschen:1990sw} and oscillons \cite{Copeland:1995fq}. As these non-perturbative calculations are distinct, the results of \cite{Lozanov:2014zfa,Hertzberg:2014jza} are likely unaffected by the issues discussed here. On the other hand, some model building considerations explored in \cite{Hertzberg:2013mba}, and certain subsequent papers, e.g.~\cite{Takeda:2014eoa,Li:2014jqa}, may need to be re-examined.

The possibility of realising baryogenesis via a complex inflaton is quite elegant, especially in its minimality. The purpose of this paper has been to compute the expected magnitude of asymmetries generated in this manner for simple models of inflation, which is a crucial step towards building more elaborate scenarios. We leave the myriad of model building opportunities for future work \cite{companion}.

\vspace{5mm}{\bf Acknowledgements}
We are grateful to  Chris Kolda for useful discussions.

\appendix
\setcounter{secnumdepth}{0}
\section{APPENDIX: CHANGING VARIABLES}
\label{AA}

In \cite{Hertzberg:2013mba} the following dimensionless EoM is considered 
\beq
\frac{\rm d^2}{{\rm d}\bar\tau^2}{\bar\rho}+3\bar H\frac{\rm d}{{\rm d}\bar\tau}{\bar\rho}+\bar\rho=0~,
\label{diml1}
\eeq
where a change of variables different to eq.~(\ref{scale}) is used, and variables with mass dimension are not scaled by a single mass scale. To see why the change of variables used in \cite{Hertzberg:2013mba} runs into difficulties, consider the general scaling 
\beq
\bar\tau\equiv M_tt~,
\qquad
\bar\rho\equiv\frac{\rho}{M_\rho}~,
\qquad
\bar H\equiv \frac{H}{M_H}~.
\label{gen}
\eeq
Then rescaling eq.~(\ref{diml1}) one obtains
\beq
\ddot{\rho}+3H\frac{M_t}{M_H}\dot{\rho}+M_t^2 \rho=0~.
\label{A3}
\eeq
Requiring that eq.~(\ref{er}) is recovered from eq.~(\ref{A3}) fixes $M_t=M_H=m$. However, this does not specify $M_{\rho}$. Moreover, as we show shortly, $M_\rho$ appears explicitly in the form of $A_\infty$ and so can not be chosen arbitrarily. In \cite{Hertzberg:2013mba} the identification $M_{\rho} =M_{\rm Pl}$ is made, causing a problem which we shall address below.

Without specifying $M_\rho$ we now rederive the form of the asymmetry $A_\infty$. Starting from eq.~(\ref{start})
\beq
A_0=
-\frac{m}{a^3\epsilon} 
\left(\frac{1}{\Lambda}\right)^{n-4} \lambda\frac{n}{2^{\frac{n}{2}-1}}\sin(\theta_i n)~I~.
\eeq
Recall from eq.~(\ref{ei}) \& (\ref{eib}) the definitions of $I$ and $\bar I$, with the scaling factor $M_\rho$ unspecified these can be related by
\beq
\bar I= \frac{m}{M_{\rho}^n} I~.
\eeq
Thus in terms of $f_n$, defined in eq.~(\ref{fn}), we obtain
\beq
A_0=
-\left(\frac{\hat \epsilon}{\epsilon}\right) \lambda f_n
\left(\frac{1}{\Lambda}\right)^{n-4} M_{\rho}^n~\sin(\theta_i n)~.
\eeq
Using $\hat \epsilon\equiv\epsilon/(m M_{\rm Pl})^2$ and replacing $f_n$ with $c_n$ we obtain
\beq
A_\infty=-\lambda c_n \frac{M_{\rho}^{n}}{m^2M_{\rm Pl}^2\Lambda^{n-4}}~\sin(\theta_i n)~.
\label{asym1}
\eeq
That the asymmetry changes with $M_\rho$ signals that this scale can not be chosen arbitrarily. Observe that taking $M_{\rho}=M_{\rm Pl}$ gives the result of \cite{Hertzberg:2013mba}, as quoted in eq.~(\ref{HK}), and for $M_{\rho}=m$ we recover eq.~(\ref{asym}), as expected.

We have already argued that an appropriate scaling amounts to a simple change of units. We next give an explicit argument in support of this approach. 
Whilst, previously we have rescaled to obtain a dimensionless EoM for $\rho$, we shall now consider the EoM for $\phi$
\beq
\ddot{\phi}+3H\dot{\phi}+m^2 \phi+ \lambda n\left(\frac{1}{\Lambda}\right)^{n-4}\phi^*{}^{n-1}=0.
\label{phi1}
\eeq 
The validity of a set of scalings is independent of $n$ and the root of the problem is most transparent for $n=4$. We consider again a general rescaling, as in eq.~(\ref{gen}) with $\bar\phi=\phi/M_\rho$. The desired form of the rescaled EoM is 
\beq
\frac{\rm d^2}{{\rm d}\bar\tau^2}{\bar\phi}+3\bar H \frac{\rm d}{{\rm d}\bar\tau}{\bar\phi}+\bar\phi+4\lambda \bar\phi^*{}^{3}=0.
\label{dimless1}
\eeq
Applying the parameter scalings we obtain
\beq
\ddot{\phi}+3H\frac{M_t}{M_H} \dot{\phi}+M_t^2\phi+4\lambda\left(\frac{M_t^2}{M_\rho^2}\right) \phi^*{}^{3}=0.
\eeq
Thus in order to recover eq.~(\ref{phi1}) with $n=4$ we require, as expected, that
\beq
m=M_\rho=M_t=M_H~.
\eeq

Finally, consider the case of general $n$, the appropriate dimensionless version of eq.~(\ref{phi1}) is
\beq
\frac{\rm d^2}{{\rm d}\tau^2}{\bar\phi}+3\hat H \frac{\rm d}{{\rm d}\tau}{\bar\phi}+\bar\phi+\lambda n\left(\frac{1}{\hat\Lambda}\right)^{n-4} \bar\phi^*{}^{n-1}=0.
\label{ffeq}
\eeq 
We rescale all dimensionful quantities (including $\Lambda$)  by a single scale $m$, except for $\phi$ which we leave unspecified, in order to show that this fixes the scaling factor for $\phi$
\beq
\tau\equiv mt~,
\quad
\bar\phi\equiv\frac{\phi}{M_\rho}~,
\quad
\hat H\equiv \frac{H}{m}~,
\quad
\hat \Lambda\equiv \frac{\Lambda}{m}~.
\label{sc}
\eeq
Applying eq.~(\ref{sc}) to eq.~(\ref{ffeq}) we obtain
\begin{equation*}
\ddot{\phi}+3H \dot{\phi}+m^2\phi+\left(\frac{m}{\Lambda}\right)^{n-4} \left(\frac{m^2}{M_\rho^{n-2}}\right)\lambda n\phi^*{}^{n-1}=0.
\end{equation*}
To recover eq.~(\ref{phi1}) we again require that $m=M_\rho$. Thus, consistent scalings should typically be in terms of a single mass scale, which is equivalent to a change of units.



\begin{thebibliography}{}
  
  

\bibitem{Sakharov:1967dj} 
  A.~D.~Sakharov,
 {\em Violation of $CP$ invariance, $C$ asymmetry, and baryon asymmetry of the universe,}
  Pisma Zh. Eksp.Teor.Fiz.{\bf 5}, 32 (1967)
  JETP Lett.{\bf 5}, 24 (1967),
  Sov.\ Phys.\ Usp.{\bf 34}, 392 (1991),
  Usp.\ Fiz.\ Nauk {\bf 161}, 61 (1991).

\bibitem{Morrissey:2012db}
  D.~E.~Morrissey and M.~J.~Ramsey-Musolf,
  {\em Electroweak baryogenesis,}
  New J.\ Phys.\  {\bf 14} (2012) 125003
  [1206.2942].
  

\bibitem{Alexander:2004us}
  S.H.S.~Alexander, M.E.~Peskin and M.M.~Sheikh-Jabbari,
 {\em Leptogenesis from gravity waves in models of inflation,}
  Phys.\ Rev.\ Lett.\  {\bf 96} (2006) 081301
  [hep-th/0403069].
  
  
\bibitem{Affleck:1984fy}
  I.~Affleck and M.~Dine,
  {\em A New Mechanism for Baryogenesis,}
  Nucl.\ Phys.\ B {\bf 249} (1985) 361.
  
\bibitem{Delepine:2006rn}
  D.~Delepine, C.~Martinez and L.~A.~Urena-Lopez,
  {\em Complex Hybrid Inflation and Baryogenesis,}
  Phys.\ Rev.\ Lett.\  {\bf 98} (2007) 161302
  [hep-ph/0609086].

\bibitem{BasteroGil:2011cx}
  M.~Bastero-Gil  {\it et al.}  
 {\em Warm baryogenesis,}
  Phys.\ Lett.\ B {\bf 712} (2012) 425
  [1110.3971].
  
\bibitem{Hertzberg:2013mba}
  M.~P.~Hertzberg and J.~Karouby,
  {\em Generating the Observed Baryon Asymmetry from the Inflaton Field,}
  Phys.\ Rev.\ D {\bf 89} (2014) 6,  063523
  [1309.0010].
  
\bibitem{Hertzberg:2013jba}
  M.~P.~Hertzberg and J.~Karouby,
  {\em Baryogenesis from the Inflaton Field,}
  Phys.\ Lett.\ B {\bf 737} (2014) 34
  [1309.0007].
  
\bibitem{Hook:2014mla}
  A.~Hook,
  {\em Baryogenesis from Hawking Radiation,}
  Phys.\ Rev.\ D {\bf 90} (2014) 8,  083535
  [1404.0113].
  
\bibitem{Hinshaw:2012aka} 
  G.~Hinshaw {\it et al.}  [WMAP Collaboration],
  {\em Nine-Year Wilkinson Microwave Anisotropy Probe (WMAP) Observations: Cosmological Parameter Results,}
  Astrophys.\ J.\ Suppl.\  {\bf 208}, 19 (2013)
  [1212.5226].

\bibitem{Ade:2015lrj}
  P.~A.~R.~Ade {\it et al.}  [Planck Collaboration],
  {\em Planck 2015. XX. Constraints on inflation,}
  [1502.02114].
  

\bibitem{Linde:1983gd}
  A.~D.~Linde,
 {\em Chaotic Inflation,}
  Phys.~Lett.~B {\bf 129} 177 (1983).
    
   
\bibitem{companion}
J.~Unwin,  in preparation.

      
\bibitem{Lozanov:2014zfa}
  K.~D.~Lozanov and M.~A.~Amin,
   {\em End of inflation, oscillons, and matter-antimatter asymmetry,}
  Phys.\ Rev.\ D {\bf 90} (2014) 8,  083528
  [1408.1811].
  

\bibitem{Hertzberg:2014jza}
  M.~P.~Hertzberg  {\it et al.}  
   {\em Theory of self-resonance after inflation. II. Quantum mechanics and particle-antiparticle asymmetry,}
  Phys.\ Rev.\ D {\bf 90} (2014) 12,  123529
  [1408.1398].
  
\bibitem{Traschen:1990sw}
  J.~H.~Traschen and R.~H.~Brandenberger,
  {\em Particle Production During Out-of-equilibrium Phase Transitions,}
  Phys.\ Rev.\ D {\bf 42} (1990) 2491.
  
\bibitem{Kofman:1994rk}
  L.~Kofman, A.~D.~Linde and A.~A.~Starobinsky,
   {\em Reheating after inflation,}
  Phys.\ Rev.\ Lett.\  {\bf 73} (1994) 3195
  [hep-th/9405187];

  
\bibitem{Copeland:1995fq}
  E.~J.~Copeland, M.~Gleiser and H.-R.~Muller,
  {\em Oscillons: Resonant configurations during bubble collapse,}
  Phys.\ Rev.\ D {\bf 52} (1995) 1920
  [hep-ph/9503217].
  
  
\bibitem{Takeda:2014eoa}
  N.~Takeda,
   {\em Inflatonic baryogenesis with large tensor mode,}
[1405.1959].
  
\bibitem{Li:2014jqa}
  N.~Li and D.~f.~Zeng,
   {\em Baryon asymmetries in a natural inflation model,}
  Phys.\ Rev.\ D {\bf 90} (2014) 12,  123542
  [1407.7386].
  

      

  \end{thebibliography}
\end{document}